# LOGISTIC MAP WITH MEMORY FROM ECONOMIC MODEL

**Valentina V. Tarasova**,

Lomonosov Moscow State University Business School, Lomonosov Moscow State University, Moscow 119991, Russia; E-mail: v.v.tarasova@mail.ru;

**Vasily E. Tarasov**,

Skobeltsyn Institute of Nuclear Physics, Lomonosov Moscow State University, Moscow 119991, Russia; E-mail: tarasov@theory.sinp.msu.ru

**Abstract.** A generalization of the economic model of logistic growth, which takes into account the effects of memory and crises, is suggested. Memory effect means that the economic factors and parameters at any given time depend not only on their values at that time, but also on their values at previous times. For the mathematical description of the memory effects, we use the theory of derivatives of non-integer order. Crises are considered as sharp splashes (bursts) of the price, which are mathematically described by the delta-functions. Using the equivalence of fractional differential equations and the Volterra integral equations, we obtain discrete maps with memory that are exact discrete analogs of fractional differential equations of economic processes. We derive logistic map with memory, its generalizations, and "economic" discrete maps with memory from the fractional differential equations, which describe the economic natural growth with competition, power-law memory and crises.

### 1. Introduction

The logistic differential equation was initially proposed in the population growth model by Verhulst [1]. In this model the rate of reproduction is directly proportional to the product of the existing population and the amount of available resources. This differential equation is actively used in economic growth models (for example, see [2, 3]). The logistic map is considered as a discrete analog of this differential equation. The logistic map, which is a simple quadratic map, demonstrates complicated dynamics, which can be characterized as universal and chaotic [4, 5, 6].

The logistic differential equation can be derived from economic model of natural growth in a competitive environment. The economic natural growth models are described by equations in which the margin output (rate of output growth) is directly proportional to income. In the description of economic growth the competition effects are taken into account by considering the price as a function of the value of output. Model of natural growth in a competitive environment is often called a model of logistic growth. We first describe the model of logistic growth, which does not take into account the memory effects.

Let Y(t) be a function that describes the value of output at time t. We assume that all manufactured products are sold (the assumption of market unsaturation). Let I(t) be a function that describes the investments made in the expansion of production, that is, the value of I(t) is the difference between the total investment and depreciation costs. In the model of natural growth, it is assumed that the marginal value of output (dY(t)/dt) is directly proportional to the value of the net investment I(t). As a result, we can use the accelerator equation

$$\frac{dY(t)}{dt} = \frac{1}{v} \cdot I(t), \qquad (1)$$

where v is a positive constant that is called the accelerator coefficient, 1/v is the marginal productivity of capital (rate of acceleration), and dY(t)/dt is the first order derivative of the function Y(t) with respect to time t.

In the logistic growth model the price P(t) is considered as a function of released product Y(t), i.e. P=P(Y(t)). The function P=P(Y) is usually considered as a decreasing function, that is, the increase of output leads to a decrease of price due to market saturation.

Assuming that the amount of net investment is a fixed part of the income P·Y(t), we get

$$I(t) = m \cdot P \cdot Y(t), \qquad (2)$$

where m is the norm of net investment (0<m<1), specifying the share of income, which is spent on the net investment.

Substituting (2) into equation (1), we obtain

$$\frac{dY(t)}{dt} = \frac{m}{v} \cdot P(Y(t)) \cdot Y(t). \qquad (3)$$

Differential equation (3) describes the economic model of natural growth in a competitive environment.

It is often assumed that the price as a function of output Y(t) is linear, i.e. P(Y(t)) =b–a·Y(t), where b is the price, which is independent of the output and a is the margin price. In this case, equation (3) has the form

$$\frac{dY(t)}{dt} = \frac{m}{v} \cdot (b - a \cdot Y(t)) \cdot Y(t). \qquad (4)$$

Equation (4) is the logistic differential equation, i.e. the ordinary differential equation of first order that describes the logistic growth. For a=0, equation (4) describes the natural growth in the absence of competition.

The logistic growth model, which is described by equation (4), and the model of the natural growth in a competitive environment, which is described by equation (3), implies that the net investment and the marginal output are connected by the accelerator equation (1). Equations (1), (3), and (4) contain only the first-order derivative with respect to time. It is known that the derivative of the first order is determined by the properties of differentiable functions of time only in infinitely small neighborhood of the point of time. As a result, the models, which are described by equations (3) and (4), assume an instantaneous change of marginal output when the net investment changes. This means not only neglecting the delay (lag) effects, but also the neglect of the memory effects, i.e. the neglect of dependence of output at the present time on the investment changes in the past. In other words, the model of logistic growth (4) does not take into account the effects of memory and delay.

## 2. Memory effect in economic process

The concept of memory is actively used in econometrics [7, 8]. We consider the concept of memory to describe economic processes by analogy with the use of this concept in physics [9, p. 394-395]. The term "memory" means the property that characterizes a dependence of the process state at a given time t=T from the process state in the past (t<T). Economic process with memory is a process, for which the economic indicators and factors (endogenous and exogenous variables) at a given time depend not only on their values at that time, but also on their values at previous time instants from a finite time interval.

A memory effect is manifested in the fact that for the same change of the economic factor, the corresponding dependent economic indicator can vary in different ways that leads us to the multivalued dependencies of indicators on factors. The multivalued dependencies are caused by the fact that the economic agents remember previous changes of this factor and indicator, and therefore can already react differently. As a result, identical changes in the present value of the factor may lead to the different dynamics of economic indicators..

To describe power-law memory we can use the theory of derivatives and integrals of non-integer order [10, 11, 12, 13]. There is an economic interpretation of the fractional derivatives [14, 15]. To take into account the effects of power-law memory, the concept of marginal values of non-integer order [16, 17] and the concept of the accelerator with memory have been proposed [18, 19]. In mathematics different types of fractional-order derivatives are known [10, 11, 12]. We will use the left-sided Caputo derivative with respect to time. One of the main distinguishing features of the Caputo fractional derivatives is that the action of these derivatives on a constant function gives zero. Using only the left-sided fractional-order derivative, we take into account the history of changes of economic indicators and factors in the past. The economic process at time t=T can depend on changes in the state of this process in the past, that is for t<T. The right-sided Caputo derivatives are defined by integration over t>T. In order to have correct dimensions of economic quantities we will use the dimensionless time variable t.

The left-sided Caputo derivative of order α>0 is defined by the formula

$$(D_{0+}^{\alpha} Y)(t) := \frac{1}{\Gamma(n-\alpha)} \int_0^t \frac{Y^{(n)}(\tau) d\tau}{(t-\tau)^{\alpha-n+1}}, \qquad (5)$$

where $\Gamma(\alpha)$ is the gamma function, $Y^{(n)}(\tau)$ is the derivative of the integer order n:=[α]+1 of the function $Y(\tau)$ with respect to the variable τ: 0<τ<t. For the existence of the expression (5), the function $Y(\tau)$ must have the integer-order derivatives up to the (n-1)th-order, which are absolutely continuous functions on the interval [0, t]. For integer orders α = n the Caputo derivatives coincide with the standard derivatives [11, p. 79], [12, p. 92-93], i.e. $(D_{0+}^{n} Y)(t) = Y^{(n)}(t)$ and $(D_{0+}^{0} Y)(t) = Y(t)$.

The generalization of the standard accelerator equation (1), which takes into account the memory effects of the order α, can be given [18] in the form

$$(D_{0+}^{\alpha} Y)(t) = \frac{1}{v} \cdot I(t), \qquad (6)$$

where v = 1 / M. For α = 1 equation (6) takes the form (1).

Note that the accelerator equation (6) includes the standard equation of the accelerator and the multiplier, as special cases [18]. This can be seen by considering equation (6) for α = 0 and α = 1. Using the property $(D_{0+}^{1} X)(t) = X^{(1)}(t)$ of the Caputo derivative, formula (6) with α = 1 takes the form of equation (1) that describes the standard accelerator. Using $(D_{0+}^{0} Y)(t) = Y(t)$, equation (6) with α = 0 is written as I(t) =v·Y(t), which is the equation of standard multiplier. Therefore, the accelerator with memory, given by equation (6), generalizes the concepts of the standard multiplier and accelerator [18].

### 3. Equation of logistic growth with memory and crises

To take into account the power-law memory effects in the natural growth model with a competitive environment, we use equation (6), which describes the relationship between the net investment and the margin output of non-integer order [16, 17]. Substituting expression (2), where P=P(Y(t)), into equation (6), we obtain

$$(D_{0+}^{\alpha} Y)(t) = \frac{m}{v} \cdot P(Y(t)) \cdot Y(t), \qquad (7)$$

where $(D_{0+}^{\alpha}Y)(t)$ is the Caputo derivative (5) of the order α≥0 of the function Y(t) with respect to time. Equation (7) is the so-called fractional differential equation with derivative of the order α> 0, [11, 12, 13]. The model of natural growth in a competitive environment, which is based on equation (7), takes into account the effects of memory with power-law fading of the order α≥0. For α = 1, equation (7) takes the form of equation (3), which describes a model of natural growth in a competitive environment without memory effects.

In the case of linearity of the price, P(Y(t))=b–a·Y (t), equation (7) has the form

$$(D_{0+}^{\alpha}Y)(t) = \frac{m}{v} \cdot (b - a \cdot Y(t)) \cdot Y(t). \qquad (8)$$

Equation (8) is the nonlinear fractional differential equation that describes the economic model of the logistic growth with memory. For α = 1, equation (8) takes the form of equation (4), which describes the logistic growth without memory effects.

If a= 0, then equation (8) takes the form of the equation of natural growth with memory

$$(D_{0+}^{\alpha}Y)(t) = \frac{m \cdot P}{v} \cdot Y(t), \qquad (9)$$

where b=P is the price, which does not depend on the value of output. Using Theorem 4.9 of [11, p. 231], we obtain the solution of equation (9) in the form

$$Y(t) = \sum_{k=0}^{n-1} Y^{(k)}(0) \cdot t^k \cdot E_{\alpha,k+1}\left(\frac{m \cdot P}{v} \cdot t^{\alpha}\right), \qquad (10)$$

where n-1<α≤n, $Y^{(k)}(0)$ is the derivative of integer order k of the function Y(t) at t=0, and $E_{\alpha,\beta}(z)$ is the two-parameter Mittag-Leffler function that is defined by the equation

$$E_{\alpha,\beta}(z) := \sum_{k=0}^{\infty} \frac{z^k}{\Gamma(\alpha k + \beta)}. \qquad (11)$$

The Mittag-Leffler function $E_{\alpha,\beta}(z)$ is a generalization of the exponential function $e^z$, such that $E_{1,1}(z) = e^z$.

If a≠0 and b≠0, we can use the variable z(t) and the parameter μ, which are defined by the equations

$$z(t) := \frac{a}{b} \cdot Y(t), \mu := \frac{m}{v}. \qquad (12)$$

Then equation of logistic growth (8) is represented in the form

$$(D_{0+}^{\alpha}z)(t) = \mu \cdot (1 - z(t)) \cdot z(t). \qquad (13)$$

This is the logistics fractional differential equation that is a fractional generalization of the logistic differential equation is proposed by Verhulst in [1]. Solution of equation is discussed in [20, 21, 22].

The crisis effects will be described as sudden changes of price in the form of price splashes (bursts) that can be represented by Gaussian functions with zero mean and small variance. It is known that the delta-function can be considered as a limit of a family of Gaussian functions with zero mean, when the variance becomes smaller [26]. For simplicity, we assume the price splashes (bursts) are periodic with period T>0 and we will describe them by the Dirac delta-function, which is a generalized function [23, 24]. The Dirac delta-function has an important role in modern economics and finance [25, 26, 27]. In general, it is possible to consider different values of the intervals between the bursts of the price. Let us consider the price function, which takes into account the periodic sharp splashes of the price, in the form

$$P(Y(t)) = -F(Y(t)) \cdot \sum_{k=1}^{\infty} \delta\left(\frac{t}{T} - k\right), \qquad (14)$$

where F(Y(t)) is a continuous function of the output Y(t) and δ(t) is the Dirac delta-function, which is a generalized function [23, 24]. The right-hand side of equation (14) makes sense if the function F(Y(t)) is continuous at the points t=kT.

### 4. Economic maps with memory and logistic maps with memory

Let us derive discrete maps with memory caused by the economic model that is described in the previous sections of this paper. Substituting expression (14) into equation (7), we get

$$(D_{0+}^{\alpha} Y)(t) = -\frac{m}{v} \cdot F(Y(t)) \cdot Y(t) \cdot \sum_{k=1}^{\infty} \delta\left(\frac{t}{T} - k\right). \qquad (15)$$

Equation (15) describes economic processes of natural growth in a competitive environment with memory and crises.

Fractional differential equation (15) contains the Dirac delta-functions, which are the generalized functions [23, 24]. The generalized functions are treated as functionals on a space of test functions. These functionals are continuous in a suitable topology on the space of test functions. Therefore equation (15) for any positive order α>0 should be considered in a generalized sense, i.e. on the space of test functions, which are continuous. In equation (15) the product of the delta-functions and the functions F(Y(t))·Y(t) is meaningful, if the function F(Y(t))·Y(t) is continuous at the points t=kT. We can use F(Y(t–ε))·Y(t–ε) with 0<ε<T (ε→0+) instead of F(Y(t))·Y(t) to make a sense of the right side of equation (15) for the case 0<α<1, when Y(kT–0)≠Y(kT+0), [49, 50, 51].

To derive discrete maps with memory from fractional differential equation (15), we use Theorem 18.19 of the monograph [9, p. 444], which is valid for any positive order α>0 and which was initially suggested in [21, 22]. The applicability of this theorem for 0<α<1 has been noted in [49, 50, 51]. Theorem 18.19 is based on the equivalence of fractional differential equations and the Volterra integral equations. Note that Lemma 2.22 of [12, p. 96–97] is the basis of this equivalence of fractional differential equations and the Volterra integral equations [12, p. 199-208]. This Lemma states that the left-sided Riemann-Liouville fractional integration provides operation, which is inverse [12, p. 96–97] to the left sided Caputo fractional differentiation that is used in equation (15). The action of the left-sided Riemann-Liouville fractional integral of the order α on equation (15) is defined on the space of test functions on the half-axis by using the adjoint operator approach [10, p.

154-157]. For fractional differential equation (15), the equivalence of fractional differential equation (15) and the Volterra integral equations should be considered in the generalized sense i.e. for the fractional differential equation with the generalized function on the space of test functions.

Using Theorem 18.19 of [9, p. 444], which is valid for any positive order α>0, we can state that the Cauchy problem with differential equation (15) and the initial conditions $Y^{(k)}(0) = Y_0^{(k)}$, where k=0, 1 ,…, N-1, and N-1<α<N, is equivalent to the following discrete map with memory

$$Y_{n+1}^{(s)} = \sum_{k=0}^{N-s-1} \frac{T^k}{k!} \cdot Y_0^{(k+s)} \cdot (n+1)^k -$$

$$\frac{m \cdot T^{\alpha-s}}{v \cdot \Gamma(\alpha-s)} \sum_{k=1}^{n}(n+1-k)^{\alpha-1-s} \cdot F(Y_k) \cdot Y_k, \qquad (16)$$

where s=0, 1, …, N-1, $Y^{(s)}(t) = d^s Y(t)/dt^s$, and

$$Y_k^{(s)} := Y^{(s)}(k \cdot T - 0) = \lim_{\varepsilon \to 0+} Y^{(s)}(k \cdot T - \varepsilon).$$

Equations (16) define a discrete map with power-law memory of the order α> 0. This map describes the natural growth in a competitive environment with memory and crises. We emphasize that discrete equation (16) is derived from the fractional differential equation (15) without the use of any approximations, i.e. it is an exact discrete analog of the fractional differential equation (15). If we will use $F(Y_k) = a \cdot Y_k - b$, then equation (16) defines the logistic map with power-law memory of the order α> 0.

For 0<α<1 (N = 1) the discrete map (16) is described by the equation

$$Y_{n+1} = Y_0 - \frac{m \cdot T^\alpha}{v \cdot \Gamma(\alpha)} \sum_{k=1}^{n}(n+1-k)^{\alpha-1} \cdot F(Y_k) \cdot Y_k, \qquad (17)$$

where n takes the positive integer values. We can write equation (17) in the form

$$Y_{n+1} = Y_0 - \frac{m \cdot T^\alpha}{v \cdot \Gamma(\alpha)} \cdot F(Y_n) \cdot Y_n - \frac{m \cdot T^\alpha}{v \cdot \Gamma(\alpha)} \sum_{k=1}^{n-1}(n+1-k)^{\alpha-1} \cdot F(Y_k) \cdot Y_k. \qquad (18)$$

Replacing n + 1 by n in equation (17), we get

$$Y_n = Y_0 - \frac{m \cdot T^\alpha}{v \cdot \Gamma(\alpha)} \sum_{k=1}^{n-1}(n-k)^{\alpha-1} \cdot F(Y_k) \cdot Y_k. \qquad (19)$$

Subtracting equation (19) from equation (18), we obtain

$$Y_{n+1} = Y_n - \frac{m \cdot T^\alpha}{v \cdot \Gamma(\alpha)} \cdot F(Y_n) \cdot Y_n - \frac{m \cdot T^\alpha}{v \cdot \Gamma(\alpha)} \sum_{k=1}^{n-1} V_\alpha(n-k) \cdot F(Y_k) \cdot Y_k, \qquad (20)$$

where $V_\alpha(z)$ is defined by $V_\alpha(z) := (z+1)^{\alpha-1} - (z)^{\alpha-1}$.

For 1<α<2 (N=2) the discrete map (16) is defined by the equation

$$Y_{n+1} = Y_0 + Y_0^{(1)} \cdot (n+1) \cdot T - \frac{m \cdot T^\alpha}{v \cdot \Gamma(\alpha)} \sum_{k=1}^{n}(n+1-k)^{\alpha-1} \cdot F(Y_k) \cdot Y_k, \qquad (21)$$

$$Y_{n+1}^{(1)} = Y_0^{(1)} - \frac{m \cdot T^{\alpha-1}}{v \cdot \Gamma(\alpha-1)} \sum_{k=1}^{n}(n+1-k)^{\alpha-2} \cdot F(Y_k) \cdot Y_k. \qquad (22)$$

For $F(Y_k) = a \cdot Y_k - b$ the discrete maps (20) and (21)-(22) describe the logistic growth with power-law memory, which the order α satisfy the condition 0<α<1 and 1<α<2 respectively.

Equations (20) and (21) - (22) describe a generalization of the logistic maps with power-law memory of the order 0<α<2.

For α=1, we can use $V_1(z) = 0$, and equation (20) gives the discrete map

$$Y_{n+1} = Y_n - \frac{m}{v} \cdot T \cdot F(Y_n) \cdot Y_n, \tag{23}$$

which describes the natural growth in a competitive environment with crises without taking into account the memory effects. Using $F(Y_k) = a \cdot Y_k - b$, equation (23) gives the logistic map

$$Y_{n+1} = \left(1 + \frac{m \cdot b \cdot T}{v}\right) \cdot Y_n - \frac{m \cdot a \cdot T}{v} \cdot Y_n^2, \tag{24}$$

which describes the logistic economic growth without the memory effects, but with the sharp splashes (bursts) of price. Equation (24) can be written as

$$Y_{n+1} = \left(1 + \frac{m \cdot b \cdot T}{v}\right) \cdot Y_n \cdot \left(1 - \frac{m \cdot a \cdot T}{v + m \cdot b \cdot T} \cdot Y_n\right). \tag{25}$$

If a≠0, we can use the variable $Z_n$ and the parameter λ, which are defined by the equations

$$Z_n := \frac{m \cdot a \cdot T}{v + m \cdot b \cdot T} \cdot Y_n, \quad \lambda := 1 + \frac{m \cdot b \cdot T}{v}. \tag{26}$$

Then equation (25) is represented in the form

$$Z_{n+1} = \lambda \cdot Z_n \cdot (1 - Z_n). \tag{27}$$

Equation (27) is the standard logistic map [4, 5, 6]. This map is used to describe different economic processes [32, 33, 34, 35]. As a result, we can state that the standard logistic map (27) can be derived from the logistic differential equation (3) without approximation only if the price function is given in the form (14), i.e. when the price behavior is described by the periodic sharp splashes (bursts) of the delta-function form.

Let us consider the logistic map with memory (20), where 0<α<1. For $F(Y_k) = a \cdot Y_k - b$, equation (20) takes the form

$$Y_{n+1} = \left(1 + \frac{m \cdot b \cdot T^\alpha}{v \cdot \Gamma(\alpha)}\right) \cdot Y_n - \frac{m \cdot a \cdot T^\alpha}{v \cdot \Gamma(\alpha)} \cdot Y_n^2 -$$

$$\frac{m \cdot T^\alpha}{v \cdot \Gamma(\alpha)} \sum_{k=1}^{n-1} V_\alpha(n-k) \cdot (a \cdot Y_k - b) \cdot Y_k. \tag{28}$$

Equation (28) can be written as

$$Y_{n+1} = \left(1 + \frac{m \cdot b \cdot T^\alpha}{v \cdot \Gamma(\alpha)}\right) \cdot Y_n \cdot \left(1 - \frac{m \cdot a \cdot T^\alpha}{v \cdot \Gamma(\alpha) + m \cdot b \cdot T^\alpha} \cdot Y_n\right) -$$

$$\frac{m \cdot b \cdot T^\alpha}{v \cdot \Gamma(\alpha)} \cdot \sum_{k=1}^{n-1} V_\alpha(n-k) \cdot Y_k \cdot \left(1 - \frac{a}{b} \cdot Y_k\right). \tag{29}$$

Using the variable $Z_n(\alpha)$ and the parameters λ(α), μ(α), ρ(α), which are defined by the expressions

$$Z_n(\alpha) := \frac{m \cdot a \cdot T^\alpha}{v \cdot \Gamma(\alpha) + m \cdot b \cdot T^\alpha} \cdot Y_n, \quad \lambda(\alpha) := 1 + \frac{m \cdot b \cdot T^\alpha}{v \cdot \Gamma(\alpha)}, \tag{30}$$

$$\mu(\alpha) := \frac{m \cdot b \cdot T^\alpha}{v \cdot \Gamma(\alpha)}, \quad \eta(\alpha) := \frac{v \cdot \Gamma(\alpha) + m \cdot b \cdot T^\alpha}{m \cdot b \cdot T^\alpha}, \tag{31}$$

we can represent equation (29) in the form

$$Z_{n+1}(\alpha) = \lambda(\alpha) \cdot Z_n(\alpha) \cdot (1 - Z_n(\alpha)) -$$
$$\mu(\alpha) \cdot \sum_{k=1}^{n-1} V_\alpha(n-k) \cdot Z_k(\alpha) \cdot (1 - \eta(\alpha) \cdot Z_k(\alpha)). \tag{32}$$

Equation (32) describes the logistic map with power-law memory of order $0<\alpha<1$. The similar form can be derived for equations (16) and (21)-(22).

The logistic map with memory (32) as well as equations (16) is an exact discrete analog of fractional differential equations (15). It should be emphasized that equations of discrete maps (16) and (32) are obtained from equation (15) without any approximations (for details, see [29, 30] and Chapter 18 of [9]).

Using (14) we can see that the logistic map with memory (32) describes a special case of economic dynamics, when price is close to zero between bursts. This behavior of price is very unusual for real economic processes. Therefore, the discrete map (32) is a toy model, but it can be used to study some properties caused by bursts of price and non-linearity.

### 5. Generalized economic and logistic maps with memory

The proposed continuous time model, which is based on equation (14), describes a very special case of economic dynamics, when price is close to zero between bursts. This behavior of price can rarely correspond to real economic processes. Therefore this model cannot be applied to real economic processes, but it can be used to describe some of their general properties. In this section, we propose a new economic model that allows us to describe the real behavior of price. These suggested models and corresponding discrete maps with memory take into account nonzero values of price between bursts of price.

Let us consider the price function, which takes into account non-zero behavior of price and the periodic sharp splashes (bursts) of price, in the form

$$P(Y(t)) = p \cdot G(Y(t)) - q \cdot F(Y(t)) \cdot \sum_{k=1}^{\infty} \delta\left(\frac{t}{T} - k\right), \tag{33}$$

where $G(Y(t))$ is a continuous function of output $Y(t)$ such that the antiderivative of the expression $(G(y) \cdot y)^{-1}$ is differentiable with respect to the variable y, and the function $F(Y(t))$ is continuous at the points $t=kT$. The parameter $q=1-p$ can be considered as the crisis measure.

For example, the function $G(Y(t))$ can be considered in the following form: (a) the constant function $G(Y(t)) = P_0$; (b) the direct proportionality $G(Y(r))=\rho \cdot Y(t)$; (c) the power-law case $G(Y(t)) = \rho \cdot Y^j(t)$. In general, the coefficients $P_0$ and $\rho$, which do not depend on $Y(t)$, are functions of time ($P_0 = P_0(t)$, $\rho = \rho(t)$).

Equation (33) generalizes the price equation (14) and the standard case without periodic sharp splashes (bursts) of price. For p=1, q=0 equation (33) corresponds to the standard case that is described by equation (3). For p=0, q=1 equation (33) takes the form (14) that corresponds to the case that is described by equation (15) with α=1.

Substituting (33) into equation (3), we obtain

$$\frac{dY(t)}{dt} = p \cdot \frac{m}{v} \cdot G(Y(t)) \cdot Y(t) - q \cdot \frac{m}{v} \cdot F(Y(t)) \cdot Y(t) \cdot \sum_{k=1}^{\infty} \delta\left(\frac{t}{T} - k\right). \tag{34}$$

We can consider the functions G(Y(t)) such that equation (34) can be represented in the form

$$\frac{dR(Y(t))}{dt} = p \cdot \frac{m}{v} \cdot C(t) - q \cdot \frac{m}{v} \cdot F_G(Y(t)) \cdot \sum_{k=1}^{\infty} \delta\left(\frac{t}{T} - k\right), \tag{35}$$

where C(t) is the function, which is independent of the output Y(t), $F_G(Y(t)) := F(Y(t))/G(Y(t))$ is the fraction of functions F(Y(t)) and G(Y(t)), and the function R(Y(t)) is defined by the equation

$$R(y) = \int_0^y (G(y) \cdot y)^{-1} dy. \tag{36}$$

Let us give simple examples of the function R(Y(t)). For example, If $G(Y(t)) = P_0$, then R(Y(t))=ln(Y(t)) and $C = P_0$; (b) If $G(Y(r)) = \rho \cdot Y(t)$, then $R(Y(t)) = -\rho \cdot Y^{-1}(t)$ and $C = \rho$; (c) If $G(Y(t)) = \rho \cdot Y^j(t)$ with $j \neq 0$, then $R(Y(t)) = -(\rho/j) \cdot Y^{-j}(t)$ and $C = \rho$.

For the economic processes with power-law memory, the generalization of the equation (35) has the form

$$(D_{0+}^{\alpha} R(Y))(t) = p \cdot \frac{m}{v} \cdot C(t) - q \cdot \frac{m}{v} \cdot F_G(Y(t)) \cdot \sum_{k=1}^{\infty} \delta\left(\frac{t}{T} - k\right), \tag{37}$$

where N-1<α<N. For 0<α<1, we can use $F_G(Y(t - \varepsilon))$ with 0<ε<T (ε→0+) instead of $F_G(Y(t))$. Let us integrate equation (37) by the Riemann-Liouville fractional integral $I_{0+}^{\alpha}$ of the order α>0 with respect to nT<t<(n+1)T. Then we get

$$(I_{0+}^{\alpha} D_{0+}^{\alpha} R(Y))(t) = p \cdot \frac{m}{v} \cdot (I_{0+}^{\alpha} C)(t) - q \cdot \frac{m}{v} \cdot I_{0+}^{\alpha} F_G(Y(t)) \cdot \sum_{k=1}^{\infty} \delta\left(\frac{t}{T} - k\right). \tag{38}$$

Using equation 2.4.42 of Lemma 2.22 of [12, p. 96], equation takes the form

$$R(Y(t)) - \sum_{k=0}^{N-1} \frac{t^k}{k!} \cdot R^k(Y(0)) = p \cdot \frac{m}{v} \cdot (I_{0+}^{\alpha} C)(t) -$$

$$q \cdot \frac{m \cdot T}{v \cdot \Gamma(\alpha)} \cdot \sum_{k=1}^{n} F_G(Y(k \cdot T)) \cdot (t - k \cdot T)^{\alpha - 1}. \tag{39}$$

Then using transformations from the proof of Theorem 18.19 of [9, p. 444] and formula 2.2.28 of [12, p. 83] in the form $(D^s I_{0+}^{\alpha} C)(t) = (I_{0+}^{\alpha-s} C)(t)$ with s<α, equation (39) gives the economic discrete map with memory

$$R_{n+1}^{(s)} = \sum_{k=0}^{N-s-1} \frac{T^k}{k!} \cdot R_0^{(k+s)} \cdot (n+1)^k + p \cdot \frac{m}{v} \cdot C_{n+1}^{(\alpha-s)} -$$

$$q \cdot \frac{m \cdot T^{\alpha-s}}{v \cdot \Gamma(\alpha-s)} \sum_{k=1}^{n} (n+1-k)^{\alpha-1-s} \cdot F_G(Y_k), \tag{40}$$

where $R^{(s)}(t) = d^s R(Y(t))/dt^s$, $R_0^{(s)} = R^{(s)}(0)$ and

$$R_k^{(s)} = R^{(s)}(k \cdot T - 0) = \lim_{\varepsilon \to 0+} R^{(s)}(k \cdot T - \varepsilon), \quad Y_k^{(s)} := Y^{(s)}(k \cdot T - 0),$$

$C_{n+1}^{(\alpha-s)} := (I_{0+}^{\alpha-s}C)((n+1)T)$, and $I_{0+}^{\alpha-s}$ is the Riemann-Liouville integration of the order α-s>0, s=0, 1, …, N-1, N-1<α<N.

For example, using Theorem 18.19 of [9, p. 444] and formula $I_{0+}^{\alpha}1 = t^{\alpha}/\Gamma(\alpha+1)$, equation (39) with constant C(t)=C gives the economic discrete map with memory in the form

$$R_{n+1}^{(s)} = \sum_{k=0}^{N-s-1} \frac{T^k}{k!} \cdot R_0^{(k+s)} \cdot (n+1)^k + p \cdot \frac{C \cdot m \cdot T^{\alpha-s}}{v \cdot \Gamma(\alpha+1-s)} \cdot (n+1)^{\alpha-s} -$$

$$q \cdot \frac{m \cdot T^{\alpha-s}}{v \cdot \Gamma(\alpha-s)} \sum_{k=1}^{n} (n+1-k)^{\alpha-1-s} \cdot F_G(Y_k), \tag{41}$$

where s=0, 1, …, N-1, N-1<α<N, $R^{(s)}(t) = d^s R(Y(t))/dt^s$, $R_0^{(s)} = R^{(s)}(0)$ and $R_k^{(s)} = R^{(s)}(k \cdot T - 0) = \lim_{\varepsilon \to 0+} R^{(s)}(k \cdot T - \varepsilon)$, $Y_k^{(s)} := Y^{(s)}(k \cdot T - 0)$.

For p=0 and q=1 equations (40) and (41) with R(Y(t))=Y(t) give the discrete map (16). The "economic" discrete maps with memory (40) and (41) describe the economic model of natural growth in a competitive environment with memory and crises effects. These discrete maps are exact discrete analogs of the fractional differential equation (37).

Using equation (41) with replacement n+1 by n, and subtraction the result from equation (41), we get the discrete map with memory of the order α>0 in the form

$$R_{n+1}^{(s)} = R_n^{(s)} + \sum_{k=0}^{N-s-1} \frac{T^k}{k!} \cdot R_0^{(k+s)} \cdot V_{k-1}(n) + p \cdot \frac{C \cdot m \cdot T^{\alpha-s}}{v \cdot \Gamma(\alpha+1-s)} \cdot V_{\alpha+1-s}(n) -$$

$$q \cdot \frac{m \cdot T^{\alpha-s}}{v \cdot \Gamma(\alpha-s)} \cdot F_G(Y_n) - q \cdot \frac{m \cdot T^{\alpha-s}}{v \cdot \Gamma(\alpha-s)} \sum_{k=1}^{n-1} V_{\alpha-s}(n-k) \cdot F_G(Y_k), \tag{42}$$

where $V_\alpha(z)$ is defined by $V_\alpha(z) := (z+1)^{\alpha-1} - (z)^{\alpha-1}$, where s=0, 1, …, N-1.

For 0<α<1 (N=1) the map (41) has the form

$$R_{n+1} = R_n + p \cdot \frac{C \cdot m \cdot T^{\alpha}}{v \cdot \Gamma(\alpha+1)} \cdot V_{\alpha+1}(n) -$$

$$q \cdot \frac{m \cdot T^{\alpha}}{v \cdot \Gamma(\alpha)} \cdot F_G(Y_n) - q \cdot \frac{m \cdot T^{\alpha}}{v \cdot \Gamma(\alpha)} \sum_{k=1}^{n-1} V_{\alpha}(n-k) \cdot F_G(Y_k). \tag{43}$$

Let us give some simple examples of the economic discrete maps with memory (41) with 0<α<1 and thus map (43). For 0<α<1 (N=1) and $G(Y(t)) = P_0$ the discrete map (41) is described by the equation

$$\ln(Y_{n+1}) = \ln(Y_0) + p \cdot \frac{P_0 \cdot m \cdot T^{\alpha}}{v \cdot \Gamma(\alpha+1)} \cdot (n+1)^{\alpha} -$$

$$q \cdot \frac{m \cdot T^{\alpha}}{v \cdot \Gamma(\alpha)} \cdot \sum_{k=1}^{n} (n+1-k)^{\alpha-1} \cdot F(Y_k). \tag{44}$$

For 0<α<1 (N = 1) and $G(Y(t)) = \rho \cdot Y^j(t)$ with j≠ 0, then economic discrete map (41) is described by the equation

$$Y_{n+1}^{-j} = Y_0^{-j} - p \cdot \frac{\rho \cdot m \cdot T^{\alpha}}{j \cdot v \cdot \Gamma(\alpha+1)} \cdot (n+1)^{\alpha} -$$

$$q \cdot \frac{m \cdot T^{\alpha}}{v \cdot \Gamma(\alpha)} \cdot \sum_{k=1}^{n} (n+1-k)^{\alpha-1} \cdot F(Y_k) \cdot Y_k^{-j}. \tag{45}$$

For j=−1 equation (45) has the form

$$Y_{n+1} = Y_0 - p \cdot \frac{\rho \cdot m \cdot T^\alpha}{j \cdot v \cdot \Gamma(\alpha+1)} \cdot (n+1)^\alpha -$$

$$q \cdot \frac{m \cdot T^\alpha}{v \cdot \Gamma(\alpha)} \cdot \sum_{k=1}^{n}(n+1-k)^{\alpha-1} \cdot F(Y_k) \cdot Y_k. \tag{46}$$

For p=0 and q=1 equation (46) gives the discrete map (17).

Discrete maps with memory (44)-(46) can be rewritten in the form similar to (43). For example, using equation (45) with replacement n+1 by n, and subtraction the result from equation (45), we get the discrete map with memory of the order 0<α<1 in the form

$$Y_{n+1}^{-j} = Y_n^{-j} - p \cdot \frac{\rho \cdot m \cdot T^\alpha}{j \cdot v \cdot \Gamma(\alpha+1)} \cdot V_{\alpha+1}(n) - q \cdot \frac{m \cdot T^\alpha}{v \cdot \Gamma(\alpha)} \cdot F(Y_n) \cdot Y_n^{-j} -$$

$$q \cdot \frac{m \cdot T^\alpha}{v \cdot \Gamma(\alpha)} \cdot \sum_{k=1}^{n-1} V_\alpha(n-k) \cdot F(Y_k) \cdot Y_k^{-j}, \tag{47}$$

where $V_\alpha(z)$ is defined by $V_\alpha(z) := (z+1)^{\alpha-1} - (z)^{\alpha-1}$.

For j=−1 equation (47) has the form

$$Y_{n+1} = Y_n - p \cdot \frac{\rho \cdot m \cdot T^\alpha}{j \cdot v \cdot \Gamma(\alpha+1)} \cdot V_{\alpha+1}(n) - q \cdot \frac{m \cdot T^\alpha}{v \cdot \Gamma(\alpha)} \cdot F(Y_n) \cdot Y_n -$$

$$q \cdot \frac{m \cdot T^\alpha}{v \cdot \Gamma(\alpha)} \cdot \sum_{k=1}^{n-1} V_\alpha(n-k) \cdot F(Y_k) \cdot Y_k. \tag{48}$$

For p=0 and q=1 equation (48) gives the discrete map (20).

Let us consider some examples of the discrete map with memory (40) with non-constant function C(t). For the power function $C(t) = C \cdot t^\beta$, where β>−1 and 0<α<1 (N=1), we can use equation 2.1.16 of [12] for the Riemann-Liouville integration $I_{0+}^\alpha t^\beta = \Gamma(\beta+1)/\Gamma(\alpha+\beta+1) \cdot t^{\beta+\alpha}$, where β>−1. For β=0, we have the equation $I_{0+}^\alpha 1 = t^\alpha/\Gamma(\alpha+1)$. Then the discrete map (40) with 0<α<1 is described by the equation

$$R_{n+1} = R_0 + p \cdot \frac{C \cdot m \cdot T^{\alpha+\beta} \cdot \Gamma(\beta+1)}{v \cdot \Gamma(\alpha+\beta+1)} \cdot (n+1)^{\alpha+\beta} -$$

$$q \cdot \frac{m \cdot T^\alpha}{v \cdot \Gamma(\alpha)} \cdot \sum_{k=1}^{n}(n+1-k)^{\alpha-1} \cdot F(Y_k). \tag{49}$$

For β=0 the map (49) takes the form (41).

If $C(t) = C \cdot t^{\beta-1} \cdot E_{\mu,\beta}(\gamma \cdot t^\mu)$, where $E_{\mu,\beta}(z) := \sum_{k=0}^\infty \frac{z^k}{\Gamma(\mu \cdot k + \beta)}$ is the two-parameter Mittag-Leffler function [12, p. 42], then we can use equation 2.2.51 [12, p. 86] to get the discrete map with memory. For example, if 0<α<1, then the discrete map with memory, which corresponds to fractional differential equation (39) with $C(t) = C \cdot t^{\beta-1} \cdot E_{\mu,\beta}(\gamma \cdot t^\mu)$, is defined by the equation

$$R_{n+1} = R_0 + p \cdot \frac{C \cdot m \cdot T^{\alpha+\beta-1}}{v \cdot \Gamma(\alpha+1)} \cdot (n+1)^{\alpha+\beta-1} \cdot E_{\mu,\alpha+\beta}(\gamma \cdot T^\mu \cdot (n+1)^\mu) -$$

$$q \cdot \frac{m \cdot T^{\alpha-s}}{v \cdot \Gamma(\alpha-s)} \sum_{k=1}^{n}(n+1-k)^{\alpha-1-s} \cdot F_G(Y_k). \tag{50}$$

Using $F(Y_k) = a \cdot Y_k - b$, equations (40)-(50) give the generalized logistic maps with memory, which describe exact discrete analogs of the economic model of generalized logistic growth in a competitive environment with memory and crises.

## 6. Conclusion

First, we briefly describe what has been suggested in this article.

1) In this paper, we proposed new economic models of the logistic growth (the natural growth in competitive environment), which take into account the power-law memory and crises. These continuous time economic models are described by fractional differential equations (15) and (37) with delta-functions.

2) Using approach, which has been proposed in works [28, 29, 30, 31], we derived exact discrete analogs of fractional differential equations (15) and (37). As a result, we got the discrete time representation of these economic models in the form of the discrete maps with memory (16) and (40).

3) We can state that the discrete maps (16), (20)-(22), and the logistic maps with memory (28), (32) are special types of the universal map with memory suggested in [28, 29, 30, 31]. In this paper, we proved that the logistic map with memory (28), (32) and the economics maps (16), (20)-(22) describe a very special case of economic dynamics, when price is close to zero between bursts. This behavior of price is very unusual for real economic processes. Therefore, this map can be considered as a toy model of real economic processes.

4) In order to have a more realistic description of the behavior of price, we proposed an economic model, which is described by equation (37), and corresponding discrete maps that are closer to real economic dynamics of price. The discrete maps with memory, which are described by equations (40) – (50), take into account nonzero values of price between bursts of price. The suggested discrete maps (40) – (50) contain the discrete maps (16), (20)-(22), and the logistic map with memory (28), (32) as special cases.

5) The suggested discrete maps with memory (40)-(50) are exact discrete analogs of the corresponding fractional differential equations (37). These maps and equations are the discrete time and continuous time representations of the economic model of the economic growth in competitive environment with memory and crises.

We now make some remarks and comments related to these results.

It should be noted that some generalizations of logistic map, which take into account memory effects, have been suggested in [36, 37, 38, 39, 40, 41, 42, 43, 44, 45]. However, these maps are not exact discrete analogs of differential equations that describe the logistic growth with memory. In this paper, we propose the discrete maps with memory as the exact discrete analogs of fractional differential equations, which describe the economic growth with competition, memory, and crises. This relationship of the fractional differential equations and the discrete maps distinguishes the suggested maps with memory from all other discrete maps with memory, which were considered in [36, 37, 38, 39, 40, 41, 42, 43, 44, 45]. The suggested discrete maps with memory are derived from the economic models, which also highlight these discrete maps.

It is known that the standard logistic map (27), which does not take into account the memory effects, can give the chaotic behavior [4, p. 33–67] and [5, 6]. Using the logistic map (27), which is derived from economic model (15) with α=1, we can state that the sudden changes of price in the form of price splashes could lead to deterministic chaotic phenomena. The suggested logistic map with memory, its generalizations and the proposed economic discrete maps with memory can demonstrate a new chaotic behavior.

The discrete maps with memory, which are exact discrete analogues of the fractional differential equations, were first proposed in works [28, 29, 30, 31]. Then, this approach, which is based on the equivalence of the fractional differential equations and the discrete maps with memory, has been applied in works [46, 47, 48, 49, 50, 51, 52, 53, 54, 55] to describe properties of the discrete maps with memory. Computer simulations of some discrete maps with memory were realized in [46, 47, 48, 49, 50, 51, 52, 53, 54, 55]. New types of chaotic behavior and new kinds of attractors have been found in these works. Therefore these types of deterministic chaotic behavior can describe some properties of the price behavior in the toy economic models, which are described by the discrete maps (16), (20)-(22), and the logistic map with memory (28), (32).

Some properties of the fractional logistic maps, which can be represented in the form (28), are investigated by computer simulation in [49, 50, 51, 52, 53, 54, 55]. In this paper, we proved that the logistic map with memory (28) and (32), and the economics maps (16) describe a very special case of economic dynamics, when price is close to zero between bursts. This behavior of price is very unusual for real economic processes. Therefore the map with memory, which is described by (16), (20)-(22), and the logistic map with memory (28), (32), can be considered only as toy models of real economic processes, but it can be used to study some properties caused by bursts of price and non-linearity.

In order to have a more realistic description of the behavior of price, we propose economic models and corresponding discrete maps (40) – (50) that are closer to real economic dynamics of price. These suggested discrete maps with memory, which are described by equations (40) – (50), take into account nonzero values of price between bursts of price. In this paper, we derive the generalized logistic map with memory and the economic discrete maps (40) – (50) from the economic model of natural growth in a competitive environment with memory and crises. The suggested discrete maps with memory are exact discrete analogs of the fractional differential equation (37) of economic dynamics. The study of properties of the generalized logistic and economic discrete maps with memory (40) – (50) requires investigations by computer simulations. The computer simulations of the suggested discrete maps with memory, which describe the natural growth in competitive environment with memory and crises, can allow us to describe new types of economic phenomena.

We should note that the fractional calculus and fractional differential equations have a wide application to describe different economic and financial processes with memory and nonlocality [56, 57, 58, 59, 60, 61, 62, 63, 64, 65, 66, 67, 68, 15, 16, 17, 18, 19] in the continuous time approach. In the framework of economic models with discrete time, the approach, which is suggested in this paper, and the approach, which is based on the exact fractional differences [69, 70, 71, 72], can be used for economic models with memory and nonlocality.

**References**

1. Verhulst P.F. Mathematical researches into the law of population growth increase // Nouveaux Mémoires de l'Académie Royale des Sciences et Belles-Lettres de Bruxelles. 1845. Vol. 18. P. 1–42. [in French]
2. Kwasnicki W. Logistic growth of the global economy and competitiveness of nations // Technological Forecasting and Social Change. 2013. Vol. 80. No. 1. P. 50–76.
3. Girdzijauskas S., Streimikiene D., Mialik A. Economic growth, capitalism and unknown economic paradoxes // Sustainability. 2012. Vol. 4. P. 2818–2837.
4. Schuster H.G. Deterministic Chaos. Fourth, Revised and Enlarged Edition. Weinheim: WILEY-VCH Verlag GmbH & Co. KGaA, 2005.
5. May R.M. Simple mathematical models with very complicated dynamics // Nature. 1976. Vol. 261 (5560). P. 459–467.
6. Baumol W., Benhabib J. Chaos: Significance, mechanism, and economic applications // Journal of Economic Perspectives. 1989. Vol.3. No.1. P.77–105.
7. Baillie R.N. Long memory processes and fractional integration in econometrics // Journal of Econometrics. 1996. Vol. 73. P. 5–59.
8. Banerjee A., Urga G. Modelling structural breaks, long memory and stock market volatility: an overview // Journal of Econometrics. 2005. Vol. 129. No. 1–2. P. 1–34.
9. Tarasov V.E. Fractional Dynamics: Applications of Fractional Calculus to Dynamics of Particles, Fields and Media. New York: Springer, 2010. 505 p.
10. Samko S.G., Kilbas A.A., Marichev O.I. Fractional Integrals and Derivatives Theory and Applications. New York: Gordon and Breach, 1993. 1006 p.
11. Podlubny I. Fractional Differential Equations. San Diego: Academic Press, 1998. 340 p.
12. Kilbas A.A., Srivastava H.M., Trujillo J.J. Theory and Applications of Fractional Differential Equations. Amsterdam: Elsevier, 2006. 540 p.
13. Diethelm K. The Analysis of Fractional Differential Equations: An Application-Oriented Exposition Using Differential Operators of Caputo Type. Berlin: Springer-Verlag, 2010. 247 p. DOI: 10.1007/978-3-642-14574-2

14. Tarasova V.V., Tarasov V.E. Economic interpretation of fractional derivatives. Progress in Fractional Differentiation and Applications. 2017. Vol. 3. No. 1. P. 1–17. DOI: 10.18576/pfda/030101
15. Tarasova V.V., Tarasov V.E. Elasticity for economic processes with memory: fractional differential calculus approach // Fractional Differential Calculus. Vol. 6. No. 2. P. 219–232. DOI: 10.7153/fdc-06-14
16. Tarasova V.V., Tarasov V.E. Marginal values of non-integer order in economic analysis // Azimuth Research: Economics and Management. 2016. No. 3 (16). P. 197–201. [in Russian]
17. Tarasova V.V., Tarasov V.E. Economic indicator that generalizes average and marginal values // Journal of Economy and Entrepreneurship. 2016. No. 11–1 (76–1). P. 817–823. [in Russian]
18. Tarasova V.V., Tarasov V.E. A generalization of concepts of accelerator and multiplier to take into account of memory effects in macroeconomics // Journal of Economy and Entrepreneurship. 2016. Vol. 10. No. 10–3. P. 1121–1129. [in Russian]
19. Tarasova V.V., Tarasov V.E. Economic accelerator with memory: Discrete time approach // Problems of Modern Science and Education. 2016. No. 36 (78). P. 37–42. DOI: 10.20861/2304-2338-2016-78-002
20. El-Sayed A.M.A., El-Mesiry A.E.M., El-Saka H.A.A. On the fractional-order logistic equation // Applied Mathematics Letters. 2007. Vol. 20. No. 7. P. 817–823.
21. West B.J. Exact solution to fractional logistic equation // Physica A: Statistical Mechanics and its Applications. 2015. Vol. 429. P. 103–108.
22. Area I., Losada J., Nieto J.J. A note on the fractional logistic equation // Physica A: Statistical Mechanics and its Applications. 2016. Vol. 444 (C). P. 182–187.
23. Lighthill M.J. Fourier analysis and generalised functions. Cambridge: Cambridge University Press, 1978.
24. Gel'fand I.M., Shilov G.E. Generalized Functions. Vol. I: Properties and Operations. Boston: Academic Press, 1964.
25. Russell T. Continuous time portfolio theory and the Schwartz-Sobolev theory of distributions // Operations Research Letters. 1988. Vol. 7. No. 3. 159–162.
26. Sato R., Ramachandran R.V. (Eds.) Conservation Laws and Symmetry: Applications to Economics and Finance. New York: Springer 1990.
27. Schulz M. Statistical Physics and Economics: Concepts, Tools, and Applications. New York: Springer-Verlag, 2003. 246 p.
28. Tarasov V.E., Zaslavsky G.M. Fractional equations of kicked systems and discrete maps // Journal of Physics A. 2008. Vol. 41. No. 43. Article ID 435101.
29. Tarasov V.E. Differential equations with fractional derivative and universal map with memory // Journal of Physics A. 2009. Vol. 42. No. 46. Article ID 465102.
30. Tarasov V.E. Discrete map with memory from fractional differential equation of arbitrary positive order // Journal of Mathematical Physics. 2009. Vol. 50. No. 12. Article ID 122703.
31. Tarasov V.E. Fractional Zaslavsky and Henon map // Long-range Interactions, Stochasticity and Fractional Dynamics. A.C.J. Luo, V. Afraimovich (Eds.). New York, Springer, 2010. P. 1–26.
32. Miskiewicz J., Ausloos M. A logistic map approach to economic cycles. (I). The best adapted companies // Physica A: Statistical Mechanics and its Applications. 2004. Vol. 336. No. 1–2. P. 206–214.

33. Ausloos M., Dirickx M. (Eds.) The Logistic Map and the Route to Chaos: From The Beginnings to Modern Applications. Part III. Berlin: Springer 2006.
34. Meyers R.A. (Ed.) Chaos and Nonlinear Dynamics: Application to Financial Markets. New York: Springer-Verlag 2011. 406 p.
35. Bischi G.I., Chiarella C., Sushko I. Global Analysis of Dynamic Models in Economics and Finance: Essays in Honour of Laura Gardini. Berlin: Springer-Verlag, 2013.
36. Fick E., Fick M., Hausmann G. Logistic equation with memory // Physical Review A. 1991. Vol. 44. No. 4. P. 2469–2473.
37. Hartwich K., Fick E. Hopf bifurcations in the logistic map with oscillating memory // Physics Letters A. 1993. Vol. 177. No. 4–5. P. 305–310.
38. Stanislavsky A.A. Long-term memory contribution as applied to the motion of discrete dynamical systems // Chaos. 2006. Vol.16. No. 4. Article ID 043105.
39. Dutta D., Bhattacharjee J.K. Period adding bifurcation in a logistic map with memory // Physica D: Nonlinear Phenomena. 2008. Vol. 237. No. 23. P. 3153–3158.
40. Alonso-Sanz R. Extending the parameter interval in the logistic map with memory // International Journal of Bifurcation and Chaos. 2011. Vol. 21. No. 1. P. 101–111.
41. Alonso-Sanz R. Discrete Systems with Memory. Singapore: World Scientific, 2011. 480 p. p. 234–245.
42. Munkhammar J. Chaos in a fractional order logistic map // Fractional Calculus and Applied Analysis. 2013. Vol. 16. No. 3. P. 511–519.
43. Wu G.C., Baleanu D. Discrete fractional logistic map and its chaos // Nonlinear Dynamics. 2014. Vol. 75. No. 1. P. 283–287.
44. Wu G.C., Baleanu D. Discrete chaos in fractional delayed logistic maps // Nonlinear Dynamics. 2015. Vol. 80. No. 4. P. 1697–1703.
45. Wu G.C., Baleanu D. Chaos synchronization of the discrete fractional logistic map // Signal Processing. 2014. Vol. 102. P.96–99.
46. Tarasov V.E., Edelman M. Fractional dissipative standard map // Chaos. 2010. Vol. 20. No. 2. (2010) Article ID 023127.
47. Edelman M., Tarasov V.E. Fractional standard map // Physics Letters A. 2009. Vol. 374. No. 2. (2009) 279–285.
48. Edelman M. Fractional standard map: Riemann-Liouville vs. Caputo // Communications in Nonlinear Science and Numerical Simulation. 2011. Vol.16. No. 12. P. 4573–4580.
49. Edelman M. Fractional maps and fractional attractors. Part I: alpha-families of maps // Discontinuity, Nonlinearity, and Complexity. 2013. Vol. 1. No. 4. P. 305–324.
50. Edelman M. Universal fractional map and cascade of bifurcations type attractors // Chaos. 2013. Vol. 23. No.3. Article ID 033127.
51. Edelman M. Fractional maps as maps with power-law memory // Nonlinear Dynamics and Complexity. Vol. 8. Edited by A. Afraimovich, A.C.J. Luo, X. Fu New York: Springer, 2014. P.79–120.
52. Edelman M. Caputo standard alpha-family of maps: fractional difference vs. fractional // Chaos. 2014. Vol. 24. No. 2. Article ID 023137.
53. Edelman M. Universality in fractional dynamics // International Conference on Fractional Differentiation and Its Applications (ICFDA), 2014. 6 p. DOI: 10.1109/ICFDA.2014.6967376 (arXiv:1401.0048).

54. Edelman M. Fractional maps and fractional attractors. Part II: Fractional difference α-families of maps // Discontinuity, Nonlinearity, and Complexity. 2015. Vol. 4. P. 391–402.
55. Edelman M. On nonlinear fractional maps: Nonlinear maps with power-law memory // Proceedings of the International Conference CCT15 – Chaos, Complexity and Transport 2015, June 1–5, 2015, Marseilles, France. (arXiv:1612.01174).
56. Scalas E., Gorenflo R., Mainardi F. Fractional calculus and continuous-time finance // Physica A: Statistical Mechanics and its Applications. 2000. Vol. 284. No. 1–4. P. 376–384.
57. Mainardi F., Raberto M., Gorenflo R., Scalas E. Fractional calculus and continuous-time finance II: The waiting-time distribution // Physica A. 2000. Vol. 287. No. 3–4. P. 468–481.
58. Laskin N. Fractional market dynamics // Physica A: Statistical Mechanics and its Applications. 2000. Vol. 287. No. 3. P. 482–492.
59. Gorenflo, R.; Mainardi, F.; Scalas, E.; Raberto, M. Fractional calculus and continuous-time finance III: the diffusion limit // In Mathematical Finance. Kohlmann A., Tang S. (Eds.) Basel: Birkhäuser, 2001. P. 171–180.
60. Cartea A., Del-Castillo-Negrete D. Fractional diffusion models of option prices in markets with jumps // Physica A. 2007. Vol. 374. No. 2. P. 749–763.
61. Vilela Mendes R. A fractional calculus interpretation of the fractional volatility model // Nonlinear Dynamics. 2009. Vol. 55. No. 4. P. 395–399.
62. Skovranek T., Podlubny I., Petras I. Modeling of the national economies in state-space: A fractional calculus approach // Economic Modelling. 2012. Vol. 29. No 4. P. 1322–1327.
63. Tenreiro Machado J., Duarte F.B., Duarte G.M. Fractional dynamics in financial indices // International Journal of Bifurcation and Chaos. 2012. Vol. 22. No. 10. Article ID 1250249. 12 p.
64. Tenreiro Machado J.A., Mata M.E. Pseudo phase plane and fractional calculus modeling of western global economic downturn // Communications in Nonlinear Science and Numerical Simulation. 2015. Vol. 22. No. 1–3. P. 396–406. DOI: 10.1016/j.cnsns.2014.08.032
65. Tejado I., Valerio D., Perez E., Valerio N. Fractional calculus in economic growth modelling: The economies of France and Italy // Proceedings of International Conference on Fractional Differentiation and its Applications. Novi Sad, Serbia, July 18 – 20. Edited by D.T. Spasic, N. Grahovac, M. Zigic, M. Rapaic, T.M. Atanackovic. 2016. P. 113–123.
66. Tarasov V.E., Tarasova V.V. Long and short memory in economics: fractional-order difference and differentiation // IRA-International Journal of Management and Social Sciences. 2016. Vol. 5. No. 2. P. 327-334. DOI: 10.21013/jmss.v5.n2.p10
67. Tarasova V.V., Tarasov V.E. Memory effects in hereditary Harrod-Domar model // Problems of Modern Science and Education. 2016. No. 32 (74). P. 38–44. DOI: 10.20861/2304-2338-2016-74-002 [in Russian]
68. Tarasova V.V., Tarasov V.E. Memory effects in hereditary Keynesian model // Problems of Modern Science and Education. 2016. No. 38 (80). P. 38–44. DOI: 10.20861/2304-2338-2016-80-001 [in Russian]
69. Tarasov V.E. Exact discrete analogs of derivatives of integer orders: Differences as infinite series // Journal of Mathematics. 2015. Vol. 2015. Article ID 134842. DOI: 10.1155/2015/134842
70. Tarasov V. E. Exact discretization by Fourier transforms // Communications in Nonlinear Science and Numerical Simulation. 2016. Vol. 37. P. 31–61. DOI: 10.1016/j.cnsns.2016.01.006

71. Tarasov V.E. Lattice fractional calculus // Applied Mathematics and Computation. 2015. Vol. 257. P. 12–33. DOI: DOI: 10.1016/j.amc.2014.11.033
72. Tarasov V.E. United lattice fractional integro-differentiation // Fractional Calculus and Applied Analysis. 2016. Vol. 19. No. 3. P. 625–664. DOI: 10.1515/fca-2016-0034